\documentclass[conference,10pt,a4paper]{IEEEtran}
\IEEEoverridecommandlockouts
\usepackage{cite}

\usepackage{makeidx,epsfig}
\usepackage{setspace,graphicx,multirow}
\usepackage{amsmath,amssymb,amsfonts}

\usepackage{subfigure}
\usepackage{graphicx}
\usepackage{epstopdf}

\usepackage{array}
\usepackage{url}
\usepackage{multirow}
\usepackage{hhline}

\usepackage{xcolor}
\usepackage{textcomp}

\usepackage{mathtools}
\usepackage{stfloats}
\usepackage{algorithm, algorithmic}
\usepackage{varwidth}

\makeatletter
\def\BState{\State\hskip-\ALG@thistlm}
\makeatother

\usepackage{fixltx2e}

\usepackage{enumitem}

\makeatletter
\newcommand\tinyv{\@setfontsize\tinyv{7pt}{9}}
\makeatother

\usepackage{authblk}

\ifodd 1
\newcommand{\com}[1]{\textbf{\color{blue} (COMMENT: #1)}} 
\else

\newcommand{\com}[1]{}
\fi

\begin{document}
\bibliographystyle{IEEEtran}
\bstctlcite{IEEEexample:BSTcontrol}
	
\title{Energy-Efficient Cyclical Trajectory Design for UAV-Aided Maritime Data Collection in Wind}
	
\author{Yifan~Zhang, Jiangbin~Lyu,~\textit{Member,~IEEE},
		and~Liqun~Fu,~\textit{Senior Member,~IEEE}
		
		
	\thanks{The authors are with the School of Informatics, and Key Laboratory of Underwater Acoustic Communication and Marine Information Technology, Xiamen University, China 361005 (email: zyfan@stu.xmu.edu.cn; \{ljb, liqun\}@xmu.edu.cn). \textit{Corresponding Author: Jiangbin Lyu.}}
}
	
\maketitle

\begin{abstract}
Unmanned aerial vehicles (UAVs), especially fixed-wing ones that withstand strong winds, have great potential for oceanic exploration and research. This paper studies a UAV-aided maritime data collection system with a fixed-wing UAV dispatched to collect data from marine buoys. We aim to minimize the UAV's energy consumption in completing the task by jointly optimizing the communication time scheduling among the buoys and the UAV's flight trajectory subject to wind effect, which is a non-convex problem and difficult to solve optimally.
Existing techniques such as the successive convex approximation (SCA) method provide efficient sub-optimal solutions for collecting small/moderate data volume, whereas the solution heavily relies on the trajectory initialization and has not explicitly considered the wind effect, while the computational complexity and resulted trajectory complexity both become prohibitive for the task with large data volume.
To this end, we propose a new cyclical trajectory design framework that can handle arbitrary data volume efficiently subject to wind effect.
Specifically, the proposed UAV trajectory comprises multiple cyclical laps, each responsible for collecting only a subset of data and thereby significantly reducing the computational/trajectory complexity, which allows searching for better trajectory initialization that fits the buoys' topology and the wind.
Numerical results show that the proposed cyclical scheme outperforms the benchmark one-flight-only scheme in general.
Moreover, the optimized cyclical 8-shape trajectory can proactively exploit the wind and achieve lower energy consumption compared with the case without wind.
\end{abstract}
	
\begin{IEEEkeywords}
	Maritime data collection, unmanned aerial vehicle, energy efficiency, wind effect, cyclical trajectory design
\end{IEEEkeywords}


\section{Introduction}
	
%
%
%
%
%
Marine areas cover almost 71\% of the Earth's surface and provide us with vast resources. New technologies to digitalize/intelligentize oceanic exploration and exploitation are fundamentally changing marine science and information network.
 More autonomous operations can increase the efficiency of managing maritime network\cite{Maritimenetworks}, while the coordinated use of various unmanned vehicles helps reducing the risk and mission costs\cite{MaritimeMissions}. Recently, unmanned aerial vehicle (UAV) finds wide applications in wireless communication systems as a mobile base station, relay or data collector (see \cite{ZengAccessing} and the references therein). Moreover, thanks to its high mobility, UAV is also a flexible and cost-effective tool that can be applied in a broad spectrum of marine activities including surveillance, rescue and data collection\cite{Buoydatacollection}.

\begin{figure}[h]
	\centering
	\setlength{\abovecaptionskip}{0.cm}
	\includegraphics[width=0.88\linewidth]{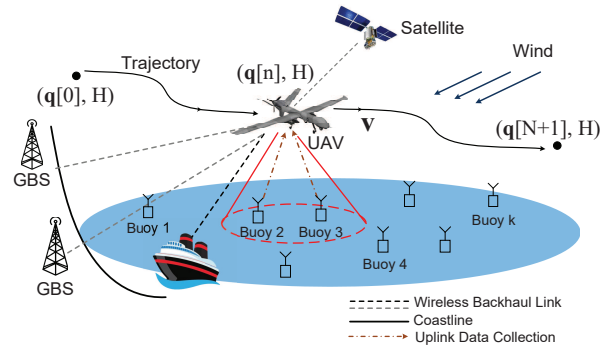}
	\caption{Maritime data collection system aided by a fixed-wing UAV in wind.\vspace{-2ex}}\label{SystemModel}
\end{figure}

\indent On the premise of collecting data from marine buoys quickly and in real-time, several telemetry activities\cite{MaritimeData} can be carried out for oceanic monitoring, research and exploitation.
Such maritime data collection can be accomplished by satellites, ships and aircrafts\cite{Roach2009Marine}, whereas satellite communication is typically costly and bandwidth-limited while manned ships/aircrafts incur high manpower/mission cost with potential risk.
In light of the above, it is particularly promising to employ a fixed-wing UAV that withstands strong winds above sea surface as an agile data collector.
Thanks to its high mobility and flexibility, UAV can fly closely to the buoys and exploit the good communication channel to wirelessly and swiftly collect large volume of data.

Although fixed-wing UAV typically can be fuel-powered and carry heavier payload than rotary-wing UAV, the limited energy onboard is still one of the critical bottlenecks for long-distance and long-endurance operations at sea. Moreover, the atmospheric drag caused by marine winds cannot be ignored, which affects the UAV's trajectory and thus restricts the UAV's flight range.
In \cite{Buoydatacollection}, UAVs are dispatched to search and recover data from buoys with optimized path planning for quality-of-service, whereas the wind effect and the UAV's energy consumption are not explicitly considered.
In \cite{NGMinimum}, the author formulates the problem of finding minimum-energy flight paths by utilizing or avoiding the wind, yet the data collection scenario is not considered and the UAV-to-buoys communication is not jointly optimized.
For energy-efficient communication design, the energy consumption models of fixed-wing \cite{ZengEnergy} or rotary-wing\cite{ZengEnergyMin} UAVs are proposed, based on which the UAV's trajectory is jointly optimized with the air-to-ground communications under various setups including data collection (e.g., \cite{UAVZhang,ZhanChengCollection}).
The problems are typically non-convex and solved sub-optimally by variants of the successive convex approximation (SCA) technique\cite{UAVZhang}.
However, these SCA-based solutions heavily rely on the trajectory initialization and have not explicitly considered the wind effect.
Moreover, for fixed-wing UAVs that must maintain a forward motion to remain aloft, the computational complexity and resulted trajectory complexity both become prohibitive for the task of collecting large volume of data from distributed buoys.

To address the above challenges, in this paper, we propose a new cyclical trajectory design framework that can handle arbitrary data volume efficiently subject to the prominent marine wind effect, which minimizes the UAV's energy consumption by jointly optimizing its trajectory and the communication time scheduling among the buoys.
Specifically, the proposed UAV trajectory comprises multiple cyclical laps, each responsible for collecting only a subset of data and hence requiring less mission time in each lap, thereby significantly reducing the computational/trajectory complexity.
Furthermore, the saved computational effort allows us to design tailored algorithms to search for better trajectory initialization that fits the buoys' topology and the wind.
Numerical results show that the proposed cyclical trajectory design outperforms the conventional one-flight-only scheme in general, in terms of energy consumption, computational time and complexity of the resulted trajectory.
Moreover, it is unveiled that the headwind that hinds the UAV's forward motion, if properly utilized, is not necessarily adversarial for the data collection task and the energy minimization.
In particular, the optimized cyclical 8-shape trajectory can proactively exploit the wind and achieve lower energy consumption compared with the case without wind, and might also outperform the simpler circular trajectory.

\section{System Model}\label{SectionModel}
As shown in Fig.\ref{SystemModel}, we consider a maritime data collection system whereby a fixed-wing UAV is dispatched as a mobile data collector to gather information from $K$ buoys on the sea surface.
Denote $\mathbf{b}_k \in {\mathbb {R}^{{\rm{2}} \times {\rm{1}}}}$ as the horizontal location of buoy $k$, $k\in\mathcal{K}\triangleq \{1,\cdots,K\}$, whose total number of information bits to be collected is denoted by ${\bar Q_{{k}}}$. Assume that the UAV has large enough data storage and can store all the collected data locally, which can be downloaded by the pilot after the UAV returns, or transferred back via the backhaul link\footnote{The backhaul link could be provided by the ground base stations (GBS) deployed along the coastline, ships equipped with high-gain antennas or the maritime satellites, as shown in Fig. \ref{SystemModel}.} if real-time data reception is required. Assume that the UAV flies at a constant altitude $H$ meters (m). Denote the UAV's trajectory projected on the horizontal plane by ${\rm{\mathbf{q}}}(t)={\big[x(t),y(t)\big]^T} \in {\mathbb {R}^{2 \times 1}}$, with $0\le t\le T$ and $T$ being the flight time. For simplicity, we dicretize the flight time $T$ into $N$+2 slots, each with sufficiently small slot time ${T_ t}$.
As a result, the UAV's trajectory is discretized as $\mathbf{q}[n]=\mathbf{q}(n{T_ t})$, $n=0,1, \ldots ,N + 1$. At any time slot $n$, the distance between the UAV and buoy $k$ is given by $d_k[n]=\sqrt {{H^2} + {{\left\| {\mathbf{q}[n] - {\mathbf{b}_k}} \right\|}^2}} $, $k\in\mathcal{K}$, with $\|\cdot\|$ denoting the Euclidean distance.

\subsection{Channel Model}
\indent Since the UAV flies at high altitude and there are no obvious obstacles at sea, we assume that the wireless channel between the UAV and buoy $k\in\mathcal{K}$ is dominated by the LoS link which follows the free-space path loss model, given by
\begin{equation}
h_k[n] = {\beta _0}{d_k[n]^{ - 2}} = \frac{{{\beta _0}}}{{{H^2} + {{\left\| {\mathbf{q}[n] - \mathbf{b}{}_k} \right\|}^2}}},
\end{equation}
where ${\beta _{\rm{0}}}$ presents the channel power gain at a reference distance of 1 m. 
Assume that each buoy transmits with power $P_0$. the achievable rate in bits/second (bps) between buoy $k$ and the UAV is expressed as
\begin{equation}
{R_k[n]} = B{\log _2}\bigg(1 + \frac{{{\gamma _0}}}{{{H^2} + {{\left\| {\mathbf{q}[n] - {\mathbf{b}_k}} \right\|}^2}}}\bigg),\label{rate}
\end{equation}
where $B$ denotes the channel bandwidth in hertz (Hz) and ${\gamma _{\rm{0}}} \triangleq {P_{\rm{0}}}{\beta _{\rm{0}}}{\rm{/}}( {{\sigma ^{\rm{2}}}\Gamma } )$ is defined as the received signal-to-noise ratio (SNR) at the reference distance of 1 m, with ${\sigma ^{\rm{2}}}$ being the noise power at the receiver and $\Gamma$$>$1 representing the channel capacity gap caused by practical modulation and coding.

\subsection{Multiple Access Scheme}
Assume that the cyclical time-division multiple access (TDMA) protocol\cite{Lyu2016Cyclical} is applied for the UAV to collect data from the $K$ buoys. At any time instant, each buoy is scheduled to communicate with the UAV only when the UAV is close enough. The UAV-enabled cyclical multiple access scheme substantially shortens the communication distance with all buoys by exploiting the mobility of the UAV, thereby improving the system throughput. For each time slot $n$, with cyclical TDMA among the $K$ buoys, denote ${\tau}_{k}[n] \ge 0$ as the allocated time for the UAV to collect data from buoy $k$. 
Then we have
\begin{equation}
\textstyle{\sum}_{k = 1}^K {\tau}_{k}[n]\le T_t^{}, \forall n.\label{allocatedtime}
\end{equation}
\indent Therefore, the total amount of information bits collected from buoy $k$ is a function of the UAV's trajectory $\{\mathbf{q}[n]\}$ and allocated time $\{{\tau}_{k}[n]\}$, which is given by\footnote{For notation simplicity, we use $\{x\}$ to denote the set of variables $x$.}
\begin{equation}
\begin{array}{l}
{Q_k}\big( {\left\{ {\mathbf{q}[n]} \right\},\left\{ {{\tau}_{k}[n]} \right\}} \big) = \sum\limits_{n = 0}^N {{\tau}_{k}[n]{R_k}[n]} \\
= B\sum\limits_{n = 0}^N {{\tau}_{k}[n]} {\log _2}\bigg(1 + \frac{{{\gamma _0}}}{{{H^2} + {{\left\| {{\mathbf{q}[n]}-{\mathbf{b}_k}} \right\|}^2}}}\bigg).
\end{array}\label{throughput}
\end{equation}
To complete the data collection from buoy $k$, we must have 
\begin{equation}
{Q_k}\big( {\left\{ {\mathbf{q}[n]} \right\},\left\{ {{\tau}_{k}[n]} \right\}} \big)\ge {\bar Q_{{k}}},\forall k \in \mathcal {K}.\label{throughputconstraint}
\end{equation}

\subsection{Energy Consumption Model for Fixed-Wing UAV}


The UAV's energy consumption mainly consists of two parts, namely the communication energy and the propulsion energy. The communication energy includes that for communication circuitry and signal transmission/reception, which is much smaller than the propulsion energy\cite{ZengEnergy} and thus ignored in this paper. For level flight, the instantaneous propulsion power required by a fixed-wing UAV with air velocity $\mathbf{v}$ and acceleration $\mathbf{a}$ is given by\cite{ZengEnergy}
\begin{equation}
P\left( {\mathbf{v},\mathbf{a}} \right)={\left| {A{{\left\| {\mathbf{v}} \right\|}^3} + \frac{C}{{\left\| {\mathbf{v}} \right\|}}\left( {1 + \frac{{{{\left\| {\mathbf{a}} \right\|}^2} - \frac{{{{({\mathbf{a}^T}\mathbf{v})}^2}}}{{{{\left\| {\mathbf{v}} \right\|}^2}}}}}{{{g^2}}}} \right)}+m{\mathbf{a}^T}\mathbf{v} \right|},\label{energyModel}
\end{equation}
where $g$ is the gravitational acceleration with nominal value 9.8 m/s$^2$, $m$ is the mass of the UAV, and $A$ and $C$ are the parameters describing the energy consumption of the UAV's movement.
It can be seen from \eqref{energyModel} that the airspeed $\mathbf{v}$ should not be too small, otherwise the required power to keep the fixed-wing UAV aloft would increase dramatically. In addition, the acceleration $\mathbf{a}$ should not be too large to cause sudden acceleration/deceleration that would consume a lot of power to generate the required thrust.

\indent Since $\mathbf{v} \triangleq \dot{\mathbf{q}}$ and $\mathbf{a} \triangleq \dot {\mathbf{v}}$ are respectively the time-varying velocity and acceleration vectors associated with the trajectory point $\mathbf{q}$ in the case of zero wind, for small time step $T_t$, we have the following results based on the first- and second-order Tayler approximations, i.e.,
\begin{equation}\label{v}
\mathbf{v}[n + 1] \approx \mathbf{v}[n] + \mathbf{a}[n]{T_t},n = 0, \ldots , N,
\end{equation}
\begin{equation}\label{q}
\mathbf{q}[n + 1] \approx \mathbf{q}[n] + \mathbf{v}[n]{T_t} + \frac{1}{2}\mathbf{a}[n]{T_t^2},n = 0, \ldots , N.
\end{equation}

As a result, with given UAV airspeed $\left\{ {\mathbf{v}[n]} \right\}$ and acceleration $\left\{ {\mathbf{a}[n]} \right\}$, the total propulsion energy is given by
\begin{equation}
\begin{array}{l}
E\big(\{ {\mathbf{v}[n]}\},\{ {\mathbf a[n]}\}\big)\approx\textstyle{\sum}_{n = 0}^{N}  P\big({\mathbf{v}[n]},\mathbf{a}[n]\big){T_t},\label{energyconsumptionn}
\end{array}\
\end{equation}
with $P( \cdot )$ given by \eqref{energyModel}.

\subsection{Wind Effect}\label{windeffect}
\begin{figure}
	\centering
	\setlength{\abovecaptionskip}{0.cm}
	\includegraphics[height=0.33\linewidth,width=0.9\linewidth]{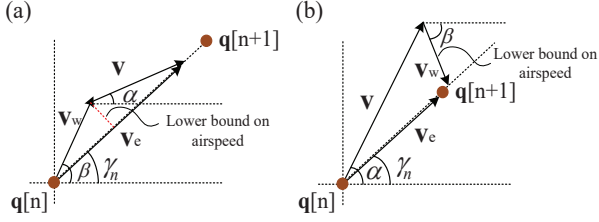}
	\caption{Illustration of (a) tailwind and (b) headwind using the wind velocity ${\mathbf{v}_w}$, as well as the UAV's airspeed ${\mathbf{v}[ n ]}$ and ground velocity ${\mathbf{v}_e[ n]}$.\vspace{-2ex}}
	\label{WindEffect}
\end{figure}

\indent The effect of wind can be regarded as a shift in the UAV's frame of reference by the wind velocity ${\mathbf{v}_w}\triangleq V_w\angle \beta$, with $V_w$ denoting its absolute value and $\beta$ denoting its angle with the positive horizontal axis, as shown in Fig. \ref{WindEffect}. For the purpose of exposition, we consider constant wind velocity in this paper. The ground velocity $\mathbf{v}_e[ n ]$ at time slot $n$ can thus be expressed in vector form as
\begin{equation}\label{vwind}
{\mathbf{v}_e[ n]} = {\mathbf{v}[ n ]} + {\mathbf{v}_w},
\end{equation}
as shown in Fig. \ref{WindEffect}, where $\alpha$ and ${\gamma _n}$ denote the angles of the UAV's airspeed and ground velocity with regard to the positive horizontal axis, respectively. Since the actual flight path is in the same frame of reference as the ground velocity, equation \eqref{q} can be re-written as
\begin{equation}\label{qinWind}
\mathbf{q}[n+1] \approx \mathbf{q}[n] + \mathbf{v}_e[n]{T_t} + \frac{1}{2}\mathbf{a}[n]{T_t^2},n = 0, \ldots , N,
\end{equation}
where the air velocity in \eqref{v} is replaced by the ground velocity.
Note that the acceleration is still controlled solely by the UAV under constant wind velocity. If the angle between the wind direction and the flight direction is less than 90$^{\circ}$, i.e., $\left| {\beta-{\gamma _n}} \right|$$<$90$^{\circ}$, the wind is called \textit{tailwind}. The tailwind promotes the UAV motion by increasing its ground velocity, so that the UAV can fly a longer distance in a time slot $T_t$. In other words, given fixed flight distance and time, the UAV only needs a lower airspeed to fly through the distance in tailwind. On the contrary, the \textit{headwind} hinders the UAV's forward motion.

\subsubsection{Effect on Communication Task}
\indent Bringing \eqref{vwind} into \eqref{qinWind}, the wind affects the UAV's ground velocity and hence its trajectory, which in turn affects the communication rates with the buoys given by \eqref{rate} and hence also the communication time scheduling among the buoys, thereby affecting the overall energy consumption given by \eqref{energyconsumptionn} in completing the data collection task given by \eqref{throughputconstraint}.
More specifically, when the UAV gets close to the buoy, the headwind reduces the ground velocity and shortens the flight distance in each time slot, allowing the UAV to maintain a good communication channel to collect data. In addition, after finishing the data collection task, the tailwind increases the ground velocity and thereby helps the UAV to reach the next trajectory point quickly. 
On the contrary, adversarial effect can happen if the UAV encounters wind of the reverse direction in the above scenarios.
Therefore, it is of great importance to jointly optimize the UAV's trajectory and communication such that the wind effect can be properly utilized without hindering the UAV's mission.

\subsubsection{Constraints on Minimum Airspeed}
There are two constraints on the UAV’s minimum airspeed. First, there is a minimum airspeed for the UAV to maintain level flight, which is known as the \textit{stall speed} and denoted by $V_s$. 
Second, the UAV needs to fly from $\mathbf{q}[n]$ to $\mathbf{q}[n+1]$ within a slot time $T_t$ subject to the wind effect.
In the case with tailwind as shown in Fig. \ref{WindEffect}(a), it is required that $||\mathbf{v}[n]|| \ge{V_w}\left| {\sin (\beta  - {\gamma _n})} \right|$.
On the other hand, in the case with headwind as shown in Fig. \ref{WindEffect}(b), it is required that $||\mathbf{v}[n]||\ge{V_w}$.

\section{Problem Formulation}

Based on the above system model, for the considered UAV-aided maritime data collection problem, we aim to minimize the UAV's energy consumption in collecting the required data volume $\bar Q_k$ from each of the $K$ buoys, by jointly optimizing the communication time scheduling among the buoys and the UAV's flight trajectory subject to wind effect.
The problem can be formulated as problem (P), where ${\mathbf{q}_{0}}$ and ${\mathbf{q}_{F}} \in {\mathbb {R}^{{\rm{2}} \times {\rm{1}}}}$ represent the UAV's initial and final locations projected onto the horizontal plane, respectively; ${V_{\max }}$ and ${a_{\max }}$ represent the maximum speed and acceleration, respectively; and ${V^{\rm{*}}}$ represents the minimum airspeed subject to the two constraints discussed in \ref{windeffect}. For simplicity, we choose ${V^{\rm{*}}}$ based on the upper bound of these two constraints, i.e., ${V^{\rm{*}}}=\max$$\{V_w, V_s\}$.
\begin{align}
\mathrm{(P)}:& \underset{
	\begin{subarray}{c}
	\{ {\mathbf{q}[n]}\} ,\{ {\mathbf{v}[n]}\},\\
	\{ {\mathbf{a}[ n ]}\},\{ {\tau_{k}[n]}\}\notag\\
	\end{subarray}
}{\min} E\big(\{ {\mathbf{v}[ n ]}\}, \{ {\mathbf{a}[ n ]}\} \big)\\	\text{s.t.}\quad & \eqref{allocatedtime}, \eqref{throughputconstraint}, \eqref{v}, \eqref{qinWind},\notag\\
&{\mathbf{q}[0]} = {\mathbf{q}_0}, {\mathbf{q}[N+1]} = {\mathbf{q}_F},\label{q0qF}\\ 
&{\mathbf{v}[0]} = {\mathbf{v}_0}, {\mathbf{v}[N+1]} = {\mathbf{v}_F},\label{v0vF}\\ 
&{V^{\rm{*}}} \le \left\| {{\mathbf{v}}[n]} \right\| \le {V_{\max }},n = 0, \ldots , N,\label{vconstraint}\\
&\left\| {\mathbf{a}[n]} \right\| \le {a_{\max }},n = 0, \ldots , N,\label{aconstraint}\\
&{\tau_{k}[n]} \ge 0,\forall k \in \mathcal {K},\forall n.\label{timeslot}
\end{align}

Problem (P) requires joint optimization of the trajectory $\{ {\mathbf{q}[n]} \}$, airspeed $\{\mathbf{v}[n]\}$, acceleration $\{\mathbf{a}[n]\}$ and communication scheduling $\{ {{\tau_{k}[n]}} \}$, which are coupled with each other through the cost function as well as the constraints \eqref{allocatedtime}, \eqref{throughputconstraint}, \eqref{v} and \eqref{qinWind}.
Moreover, due to the non-convex throughput constraint \eqref{throughputconstraint} and the non-convex cost function $E\big(\{ {\mathbf{v}[ n ]}\}, \{ {\mathbf{a}[ n ]}\} \big)$ for the fixed-wing UAV's energy consumption, problem (P) is further complicated and cannot be directly solved using the standard convex optimization techniques.

\section{Proposed Solution}\label{ProposedSolution}

To solve the complicated problem (P), we leverage the SCA technique as the basic optimization tool, which approximates each non-convex function involved in (P) by a convex and differentiable function based on the first-order Taylor approximation at a certain local point\cite{ZengEnergy}. A sub-optimal solution to the original non-convex problem can then be obtained by solving a series of convex sub-problems with successively updated local points at each iteration. The computational complexity of each convex sub-problem can be shown to grow in the order of $O({N^{3.5}})$ based on similar analysis given in \cite{UAVZhang}, where $N$ is the number of time slots required to complete the data collection task in our considered setup. 

Such a SCA-based solution can be obtained efficiently for small/moderate data volume to be collected, which corresponds to short mission time required and hence small number of time slots $N$ in completing the task.
However, in our considered maritime data collection scenario, it is likely that each buoy stores a large volume of historical maritime/undersea monitoring data, which requires prolonged mission time for the UAV to complete the data collection.
As a result, for the fixed-wing UAV that must maintain a forward motion to remain aloft, the computational complexity and resulted trajectory complexity both become prohibitive for the task of collecting large volume of data from distributed buoys.
In addition, the SCA-based solution heavily relies on the trajectory initialization and may trap in some locally optimal point, whereby it may need to try out several different initial trajectories in order to approach the globally optimal solution.
Moreover, the effect of wind on the data collection task has not been explicitly considered in the literature, which in fact could be exploited to design tailored trajectories for minimizing the UAV's energy consumption in completing the data collection task.

To address the above challenges, we propose a new cyclical trajectory design framework that can handle arbitrary data volume efficiently subject to the prominent marine wind effect.
Specifically, the proposed UAV trajectory comprises $M\geq 1$ cyclical laps, each responsible for collecting only a fraction $1/M$ of data and hence requiring less mission time in each lap, thereby significantly reducing the computational/trajectory complexity.
Furthermore, the saved computational effort allows us to design tailored algorithms to search for better trajectory initialization that fits the buoys' topology and the wind.
For the purpose of exposition, we consider two simple patterns of initial trajectories, namely the circular trajectory and the 8-shape trajectory, and assume for simplicity that each buoy has the same data volume $Q$ to be collected.
The detailed algorithms are summarized in Algorithms 1 and 2.

\begin{algorithm}\caption{Cyclical Trajectory Optimization}\label{CyclicalTrajectory}
	\begin{small}
		\textbf{Input:} Buoy locations $\mathbf{b}_k, k\in\mathcal{K}$, wind velocity ${\mathbf{v}_w}$, data volume $Q$ and the number of laps $M$.\\
		\textbf{Output:} Cyclical trajectory $\{{\mathbf{q}[n]}^{\rm{*}}\}$, airspeed $\{{\mathbf{v}[n]}^{\rm{*}}\}$, acceleration $\{{\mathbf{a}[n]}^{\rm{*}}\}$ and communication scheduling $\left\{ {{\tau_{k}[n]^*}} \right\}$.
		
		\begin{algorithmic}[1]
			\STATE Set $Q_0$ = ${Q/M}$ and initialize the maximum number of iterations $l_0$.
			\STATE Choose one pattern of initial trajectory (e.g., circular or 8-shape).
			\STATE Call \textbf{InitialTrajectory}$\big(Q_0, \{\mathbf{b}_k\}, \mathbf{v}_w\big)$ and obtain the time period $T_0$ and initialization $\{\mathbf{q}[n]^0\}$, $\{\mathbf{v}[n]^0\}$ and $\left\{ {{\tau_{k}[n]^0}} \right\}$.\label{StepInitial}
			\STATE Solve problem (P) based on SCA and obtain the solution $\{{\mathbf{q}[n]}\}$, $\{{\mathbf{v}[n]}\}$, $\{{\mathbf{a}[n]}\}$ and $\left\{ {{\tau_{k}[n]}} \right\}$. Record the energy consumption $E$.\label{StepSCA}
			\REPEAT\label{Alg1Repeat}
			\STATE Fine-tune the time period $T$ around $T_0$, with corresponding modification on the initialization obtained in Step \ref{StepInitial}.
			\STATE Call step \ref{StepSCA}.
			\UNTIL The optimal solution is found or $l_0$ is reached. 
			\STATE Output the solution with the minimum $E$ recorded.\label{Alg1output}
		\end{algorithmic}
	\end{small}
\end{algorithm}

Note that for large data volume, the communication time that the UAV enters/leaves the cyclical trajectory can be practically ignored.
Our proposed cyclical trajectory optimization in Algorithm \ref{CyclicalTrajectory} consists of two main phases, i.e., the initialization phase (steps 1$\sim$\ref{StepSCA}) and the fine-tuning phase (steps \ref{Alg1Repeat}$\sim$\ref{Alg1output}).
The initialization phase efficiently finds a simple feasible solution of problem (P) by fixing the trajectory pattern (e.g., circular trajectory with one circle of radius $r$ centered at the origin, or 8-shape trajectory with two circles of radius $r$ tangent to each other).
Given the trajectory pattern, the \textbf{InitialTrajectory} procedure searches for the optimal time period $T_0$ and trajectory parameters (e.g., the radius $r$ and/or the orientation $\theta$ of the 8-shape trajectory, illustrated later in Fig. \ref{8-shape}(a)) that achieve the least possible energy consumption within a certain maximum number of iterations (denoted by $l_1$, $l_2$, etc.) while satisfying the throughput constraint.
Note that the throughput feasibility test under a given trajectory pattern only involves simple linear inequalities with the time allocation $\left\{ {{\tau_{k}[n]}} \right\}$, which can be done much faster than solving the original problem (P) using SCA.
Based on the obtained feasible time period $T_0$ and initialization $\{\mathbf{q}[n]^0\}$, $\{\mathbf{v}[n]^0\}$ and $\left\{ {{\tau_{k}[n]^0}} \right\}$, we can then fine-tune the time period $T$ around $T_0$ (for a maximum of $l_0$ iterations), and apply SCA in each iteration to fine-tune the trajectory and time allocation, hence further reducing the energy consumption but with much fewer calls of the SCA routine.

Finally, note that by partitioning into $M$ laps, 
the mission time and hence the number of time slots in each lap can be roughly reduced to $1/M$ of the one-flight-only scheme, whereby the computational complexity can be reduced to around ${1/O({M^{3.5}})}$ when SCA is applied.
Furthermore, the saved computational effort allows us to search for better trajectory initialization that fits the buoys' topology and the wind, which is reduced to feasibility tests on linear inequalities and hence involves much lower complexity than SCA.
Moreover, thanks to the reduced time period $T$ and the simpler cyclical trajectory, it is typically much easier to search and fine-tune the mission time and trajectory initialization\footnote{Note that these steps are also needed in the one-flight-only scheme.} before feeding to the SCA routine.
Therefore, our proposed cyclical trajectory design framework can efficiently reduce the UAV's energy consumption in completing the data collection task, especially for large data volume to be collected.

\begin{algorithm}\caption{\textbf{InitialTrajectory} Procedure}\label{InitialTrajectory}
	\begin{small}
		$\big[T_0, \{\mathbf{q}[n]^0\}, \{\mathbf{v}[n]^0\}, \left\{ {{\tau_{k}[n]^0}} \right\} \big]$=\textbf{InitialTrajectory}$\big(Q_0, \{\mathbf{b}_k\}, \mathbf{v}_w\big)$
		
		\begin{algorithmic}[1]
			\STATE Assume constant ground speed $V$ (0 $\le$ $V$ $\le$ $V_\textrm{max}$). Set the geometric center of $\{\mathbf{b}_k\}$ as the origin. Initialize $l_1$, $l_2$ and $l_3$.
			\REPEAT 
			\STATE Search for $T_0$.\label{SearchT0}
			\REPEAT 
			\STATE Search for $r$. Obtain $V$ based on 2$\pi r=VT$ for circular trajectory and 2$\pi r$ = ${VT/2}$ for 8-shape trajectory.\label{Searchr}
			\STATE (For 8-shape trajectory: search for optimal orientation $\theta$.)
			\STATE Given $T_0$, $r$, $V$ and $\mathbf{v}_w$, obtain the energy consumption. Find feasible $\left\{ {{\tau_{k}[n]^0}} \right\}$ by solving (P) via linear programming. 
			\STATE If it is infeasible, return to step \ref{Searchr}.
			\UNTIL The optimal solution is found or $l_2$ is reached.
			\STATE If it is infeasible, return to step \ref{SearchT0}.
			\UNTIL $T_0$ that minimizes the energy is found or $l_1$ is reached.
			\STATE Given $T_0$, $r$, $V$ and $\mathbf{v}_w$, obtain $\{\mathbf{q}[n]^0\}$, $\{\mathbf{v}[n]^0\}$ and $\left\{ {{\tau_{k}[n]^0}} \right\}$.	
		\end{algorithmic}
	\end{small}
\end{algorithm}

\section{Numerical Results}
This section provides numerical results to validate the proposed design. The following parameters are used if not mentioned otherwise: $H=100$ m, $B=1$ MHz, ${\gamma _{\rm{0}}}=70$ dB, $V_\textrm{max}=100$ m/s, $V_s=3$ m/s, $a_\textrm{max}=5$ m/s$^2$, $A=9.26 \times {10^{ - 4}}$ and $C=2250$ \cite{ZengEnergy}.
For the purpose of exposition, we first consider the single-buoy scenario and investigate two cases including the chain-like flight (explained later) and cyclical flight, which help to illustrate the effect of wind on the data collection task, as well as the impact of data volume on the trajectory complexity and resulted energy consumption.
Then we extend to the multi-buoy case, and investigate the situations where the distributed buoys are far from/close to each other.

\subsection{Chain-Like Flight (Single Buoy)}\label{SectionChain}
\begin{figure}[h]
	\centering
	\vspace{-2ex}
	\setlength{\abovecaptionskip}{0.cm}
	\includegraphics[height=0.52\linewidth,width=0.73\linewidth]{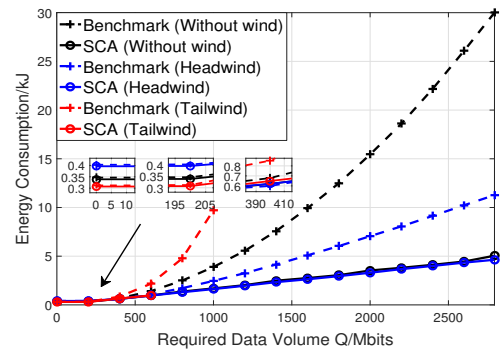}
	\caption{Energy consumption versus data volume requirement under different wind conditions.\vspace{-1ex}}
	\label{QVE}
\end{figure}

Assume $\mathbf{q}_{0}={[-600,0]^T}$ m and $\mathbf{q}_{F}={[600,0]^T}$ m with the single buoy at the origin. In this case, the initial point $\mathbf{q}_{0}$, the buoy and the final point $\mathbf{q}_{F}$ make up a chain-like topology, and hence the UAV's flight is termed as the \textit{chain-like flight}.
The UAV's initial trajectory is set to be the direct path from $\mathbf{q}_\textrm{0}$ to $\mathbf{q}_{F}$.
Assume that the wind speed is $V_w=5$ m/s, which either blows from $\mathbf{q}_{0}$ to $\mathbf{q}_{F}$ (i.e., tailwind) or reversely (i.e., headwind).
We use SCA to optimize the UAV's trajectory and airspeed, and also compare with the benchmark scheme where the UAV flies along a straight line at a constant but optimized airspeed. 
The resulted energy consumption for collecting different data volume $Q$ is shown in Fig. \ref{QVE}, under different wind conditions.

For the benchmark scheme,
in the case with small $Q$ (e.g., $Q\leq 200$ Mbits), it is observed that the tailwind helps achieve lower energy consumption compared with the headwind or no-wind case.
This is because collecting small data volume $Q$ can be done easily, and the main task of the UAV is to fly from $\mathbf{q}_{0}$ to $\mathbf{q}_{F}$, for which the tailwind helps.
On the other hand, as $Q$ increases (e.g., $Q\geq 400$ Mbits), the tailwind consumes the most energy while the headwind consumes the least energy, which might be \textit{against the intuition} that the tailwind adds thrust to the UAV and hence should help save energy.
The underlying reason is that, to collect more data, the UAV has to slow down to allow more time to communicate with the buoy, and hence might not fly at the most energy-saving airspeed\footnote{Based on \eqref{energyModel}, there exists a most energy-saving airspeed, since the UAV consumes much energy at both high airspeed and also low airspeed (in order to keep the UAV aloft).}.
Note that due to the limited communication time, the UAV might not be able to collect too large data volume (e.g., $Q> 1000$ Mbits) in the tailwind case.

\begin{figure}[h]
	\centering
	\vspace{-2ex}
	\setlength{\abovecaptionskip}{0.cm}
	{
		\label{a} 
		\includegraphics[height=0.36\linewidth,width=0.49\linewidth]{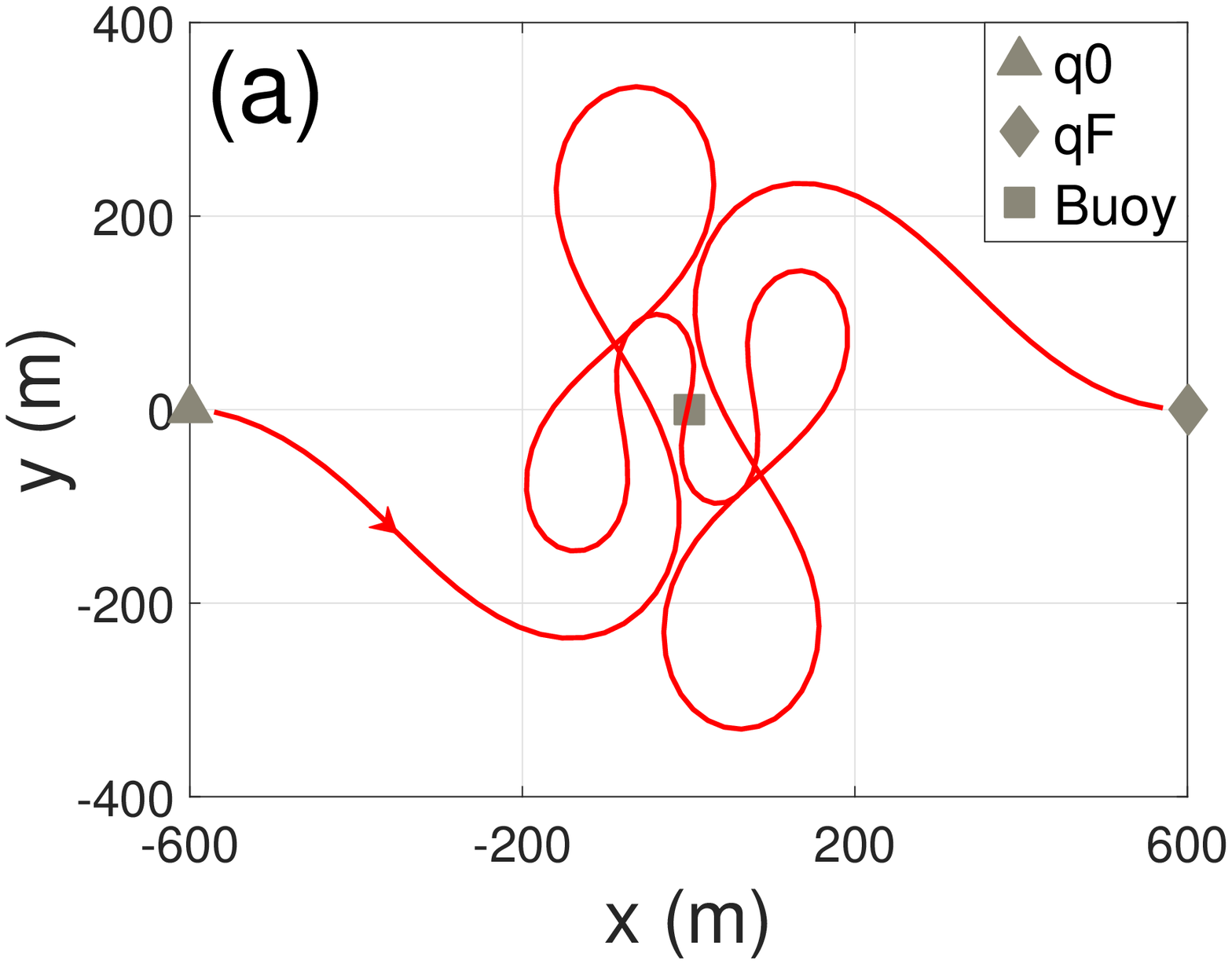}}
	{
		\label{b} 
		\includegraphics[height=0.36\linewidth,width=0.47\linewidth]{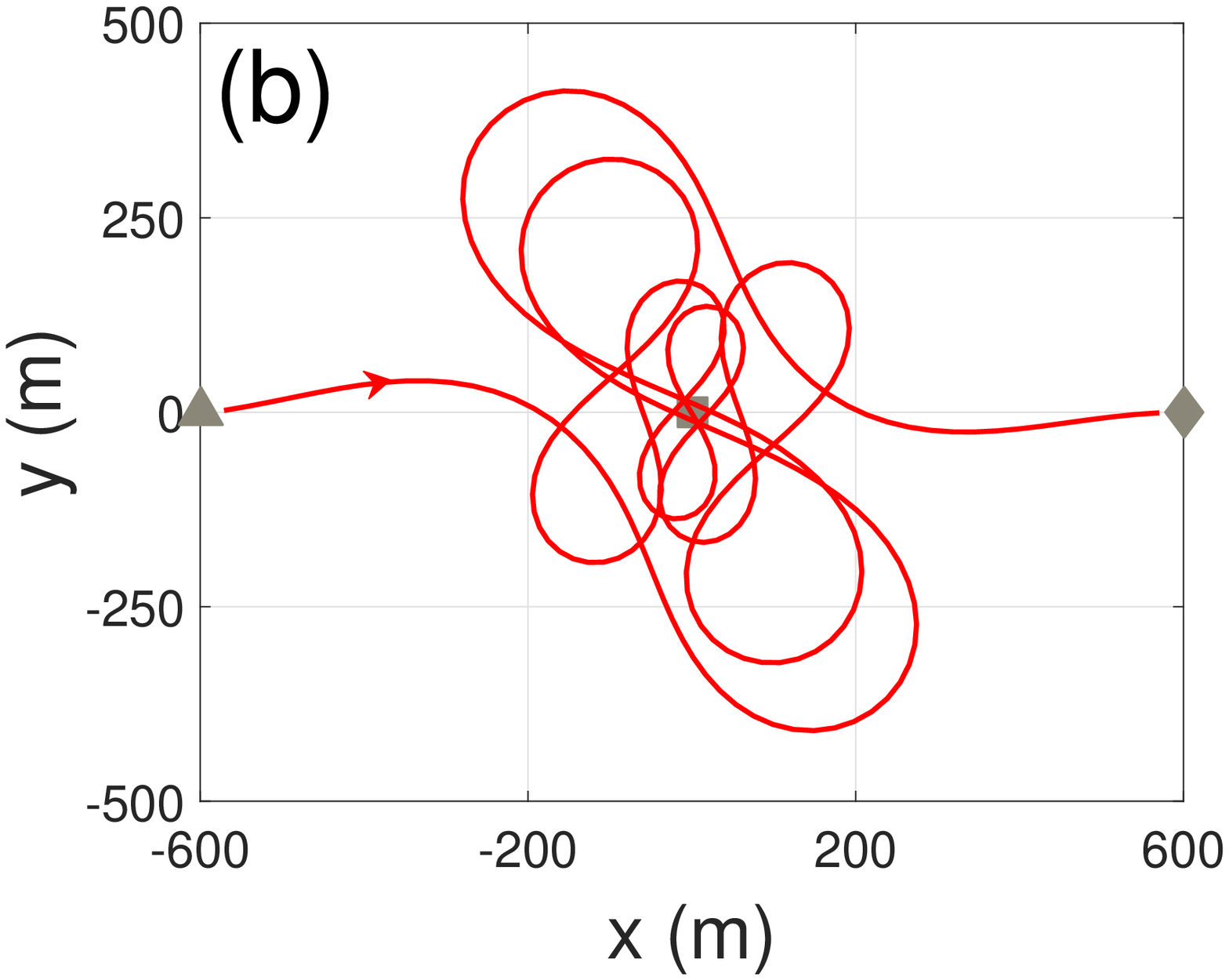}}
	\caption{The SCA-based UAV trajectory for the chain-like flight without wind, under (a) $Q=1400$ Mbits and (b) $Q=2200$ Mbits.\vspace{-1ex}}
	\label{CurveTrajectory} 
\end{figure}

In comparison, regardless of the wind conditions, our proposed SCA-based scheme is able to adapt to the wind and achieve significant energy savings compared with the benchmark scheme.
However, it is worth noting that as the required data volume increases, the computational complexity and the resulted trajectory complexity of the SCA-based solution both increase dramatically.
Two examples of the UAV trajectory under different $Q$ are shown in Fig. \ref{CurveTrajectory}. 
This thus motivates our cyclical trajectory design to achieve lower energy consumption with less computational time and simpler UAV trajectory.

\vspace{-2ex}
\subsection{Cyclical Trajectory (Single Buoy)}
\begin{figure}[h]
	\centering
	\setlength{\abovecaptionskip}{0.cm}
	\includegraphics[height=0.53\linewidth,width=0.73\linewidth]{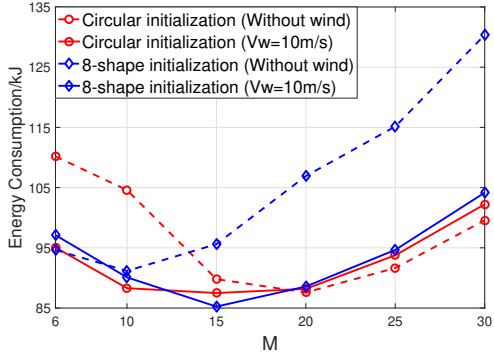}
	\caption{Total energy consumption of all $M$ laps under different trajectory initializations, without wind or with wind.\vspace{-1ex}}
	\label{MVE}
\end{figure}

Consider a large data volume, e.g., $Q=6$ Gbits.
For our considered single-buoy setup, the proper amount of data volume $Q_0=Q/M$ for each lap should be around 200 to 1000 Mbits, based on the observations in Section \ref{SectionChain}.
Therefore, we choose the number of laps $M$ in the range of 6 to 30. 
Based on our propose cyclical trajectory optimization in Algorithm \ref{CyclicalTrajectory}, the resulted total energy consumption of all $M$ laps is shown in Fig. \ref{MVE}, under different trajectory initializations and wind conditions.
It is observed that with our proposed cyclical trajectory optimization, both the optimized circular and 8-shape trajectories can proactively exploit the wind to reduce the energy consumption under a certain $Q_0$ per lap, compared with the case without wind.
In particular, the 8-shape trajectory may even make better use of the wind and outperform the circular trajectory in some cases (e.g., under $Q_0=400$ Mbits and $V_w=10$ m/s).
We provide more detailed discussions next.

\subsubsection{Without Wind}
In this case, the circular trajectory performs better than the 8-shape trajectory when $M\geq 15$, where the minimum total energy consumption occurs at $M=20$ (i.e., $Q_0=300$ Mbits) with the circular trajectory. 
This is because when $Q_0$ is relatively small, the UAV can collect data by flying circularly with a moderate acceleration, without incurring much turning energy as in the 8-shape trajectory.
On the other hand, when $Q_0$ is large, it is more beneficial to adopt the 8-shape flight whose trajectory points are overall closer to the buoy and hence enjoy better communication channel.

\subsubsection{With Wind}
\begin{figure}[h]
	\centering
	\vspace{-2ex}
	\setlength{\abovecaptionskip}{0.cm}
	{
		\includegraphics[height=0.375\linewidth,width=0.5\linewidth]{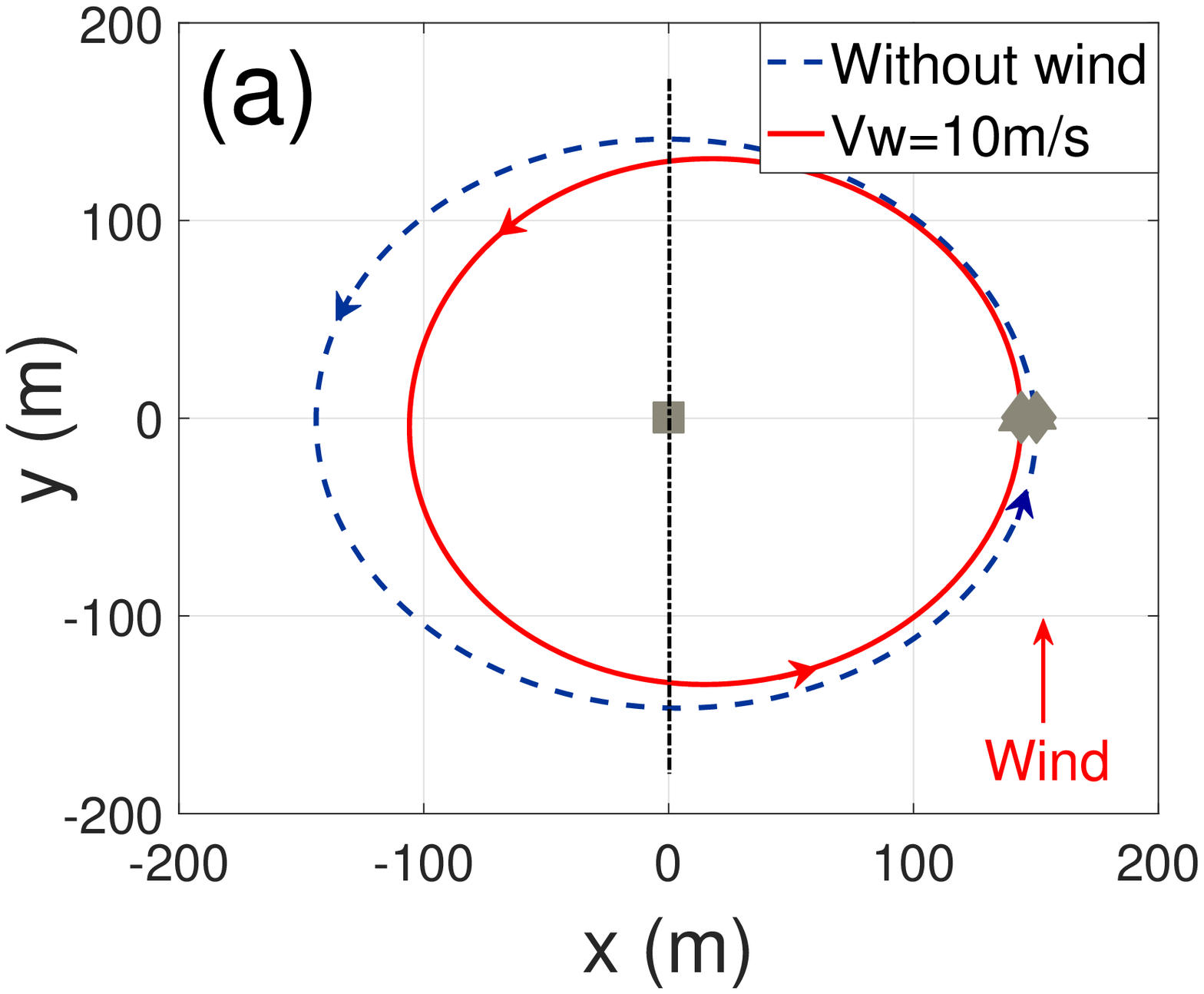}}
	{
		\includegraphics[height=0.375\linewidth,width=0.475\linewidth]{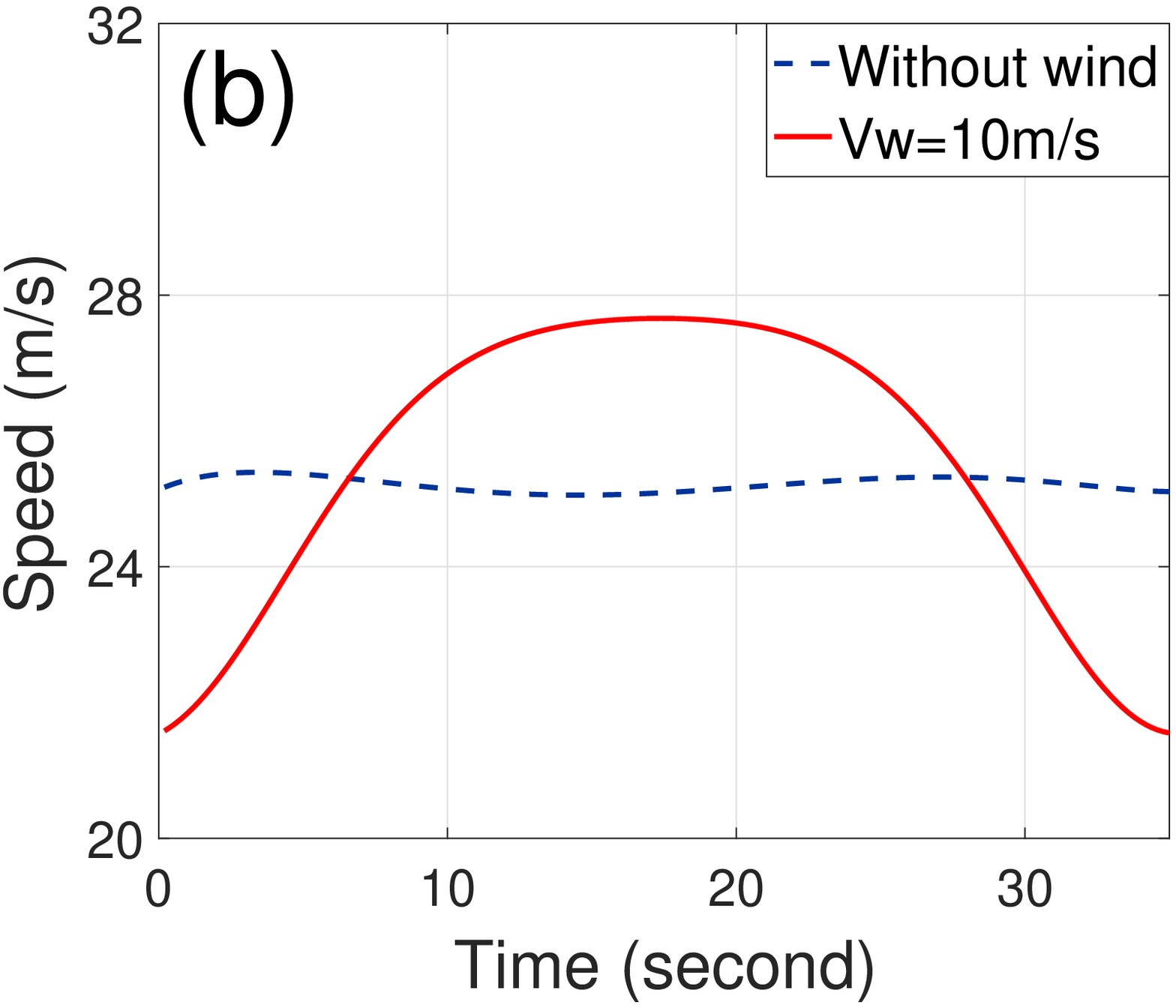}}
	\caption{Optimized UAV (a) trajectory and (b) airspeed for circular trajectory.\vspace{-1ex}}
	\label{Circular} 
\end{figure}

\begin{figure}[h]
	\centering
	\vspace{-2ex}
	\setlength{\abovecaptionskip}{0.cm}
	{
		\includegraphics[height=0.375\linewidth,width=0.5\linewidth]{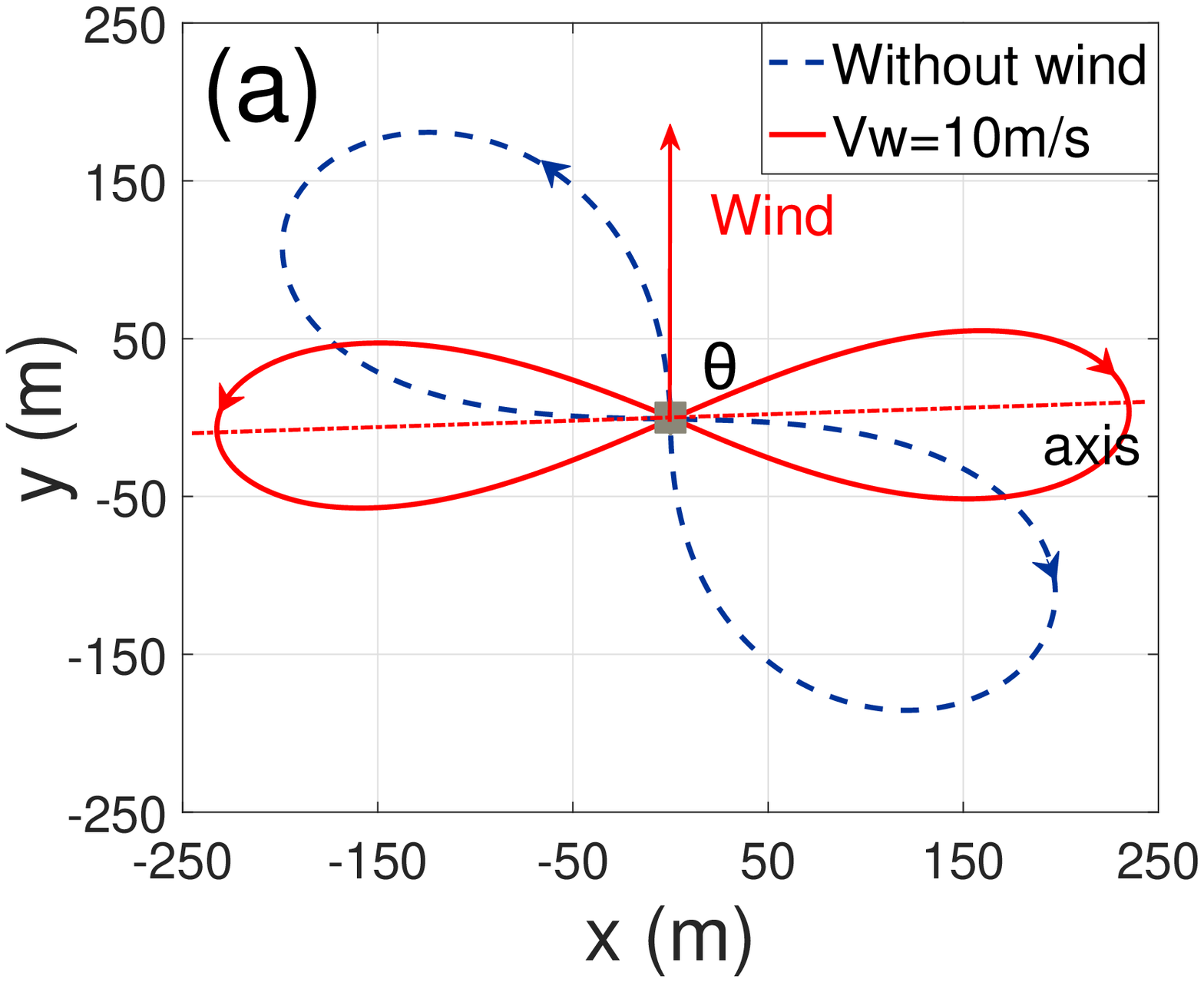}}
	{
		\includegraphics[height=0.375\linewidth,width=0.475\linewidth]{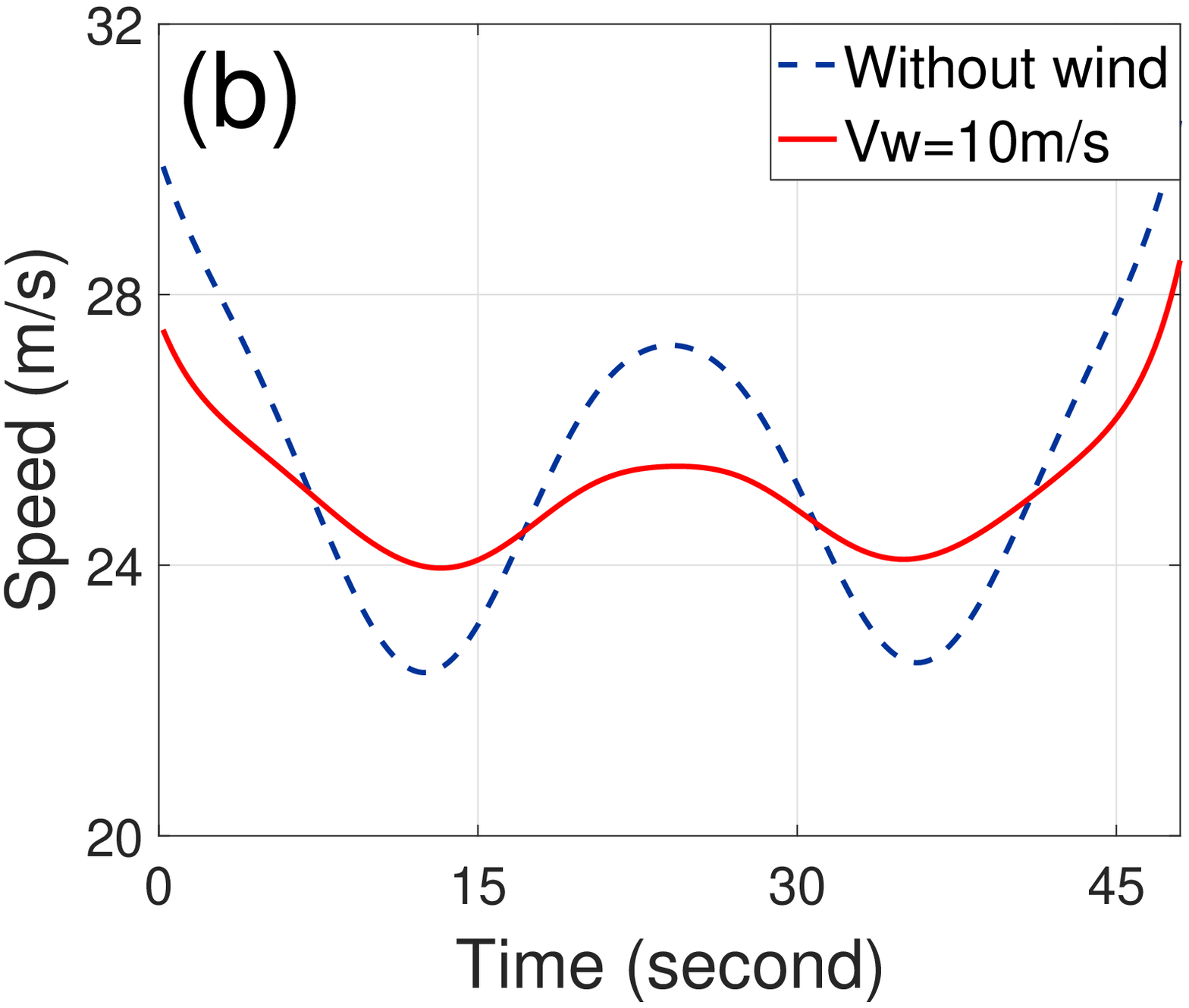}}
	\caption{Optimized UAV (a) trajectory and (b) airspeed for 8-shape trajectory.\vspace{-1ex}}
	\label{8-shape} 
\end{figure}

Consider wind blowing from south to north. The optimized UAV trajectory and airspeed for the circular and 8-shape trajectories are shown in Fig. \ref{Circular} and Fig. \ref{8-shape}, respectively.
The circular trajectory is divided into two halves, where the left half experiences headwind and the right half experiences tailwind. The trajectory optimization needs to balance between the UAV's airspeed and angular acceleration in achieving lower energy consumption subject to wind effect.
As for the 8-shape trajectory, it becomes more flatten in wind in order to reduce the overall distance to the buoy.
Moreover, the optimized \textit{orientation} $\theta$, i.e., the angle between the wind and the axis of the 8-shape, is around $90^{\circ}$ as shown in Fig. \ref{8-shape}(a), and hence the UAV experiences headwind in both ends of the 8-shape, whereby the UAV can exploit the wind to slow down and hence reduce the turning energy.

Finally, it is worth noting that the recommended data volume per lap (e.g, $Q_0=300$ Mbits using the optimized circular trajectory for the case without wind, or $Q_0=400$ Mbits using the optimized 8-shape trajectory for the case with wind) serves as a good reference for partitioning arbitrarily large data volume $Q$ into $M=\lceil Q/Q_0\rceil$ laps.

\subsection{Multi-buoy Case}
\subsubsection{Long Buoy Distance}
In this case, the buoys are distributed in a large area and are far from each other.
If the data volume $Q$ to be collected from each buoy is large, we could devise a proper visiting order of the buoys, and then apply our cyclical trajectory design for each one of them.

\begin{figure}[h]
	\centering
	\vspace{-2ex}
	\setlength{\abovecaptionskip}{0.cm}
	{
		\label{MTrajectory} 
		\includegraphics[height=0.36\linewidth,width=0.50\linewidth]{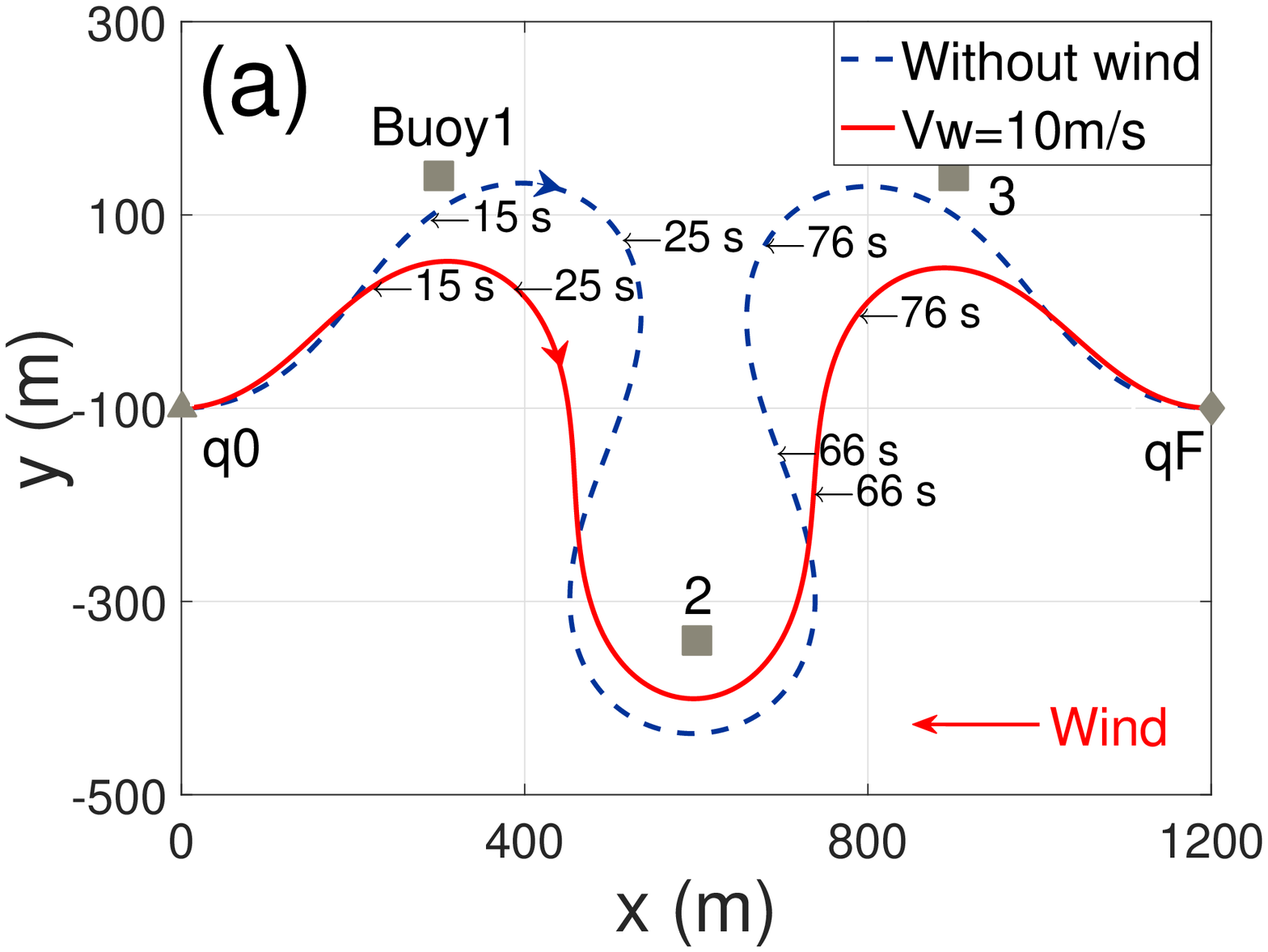}}
	{
		\label{TDMA} 
		\includegraphics[height=0.372\linewidth,width=0.465\linewidth]{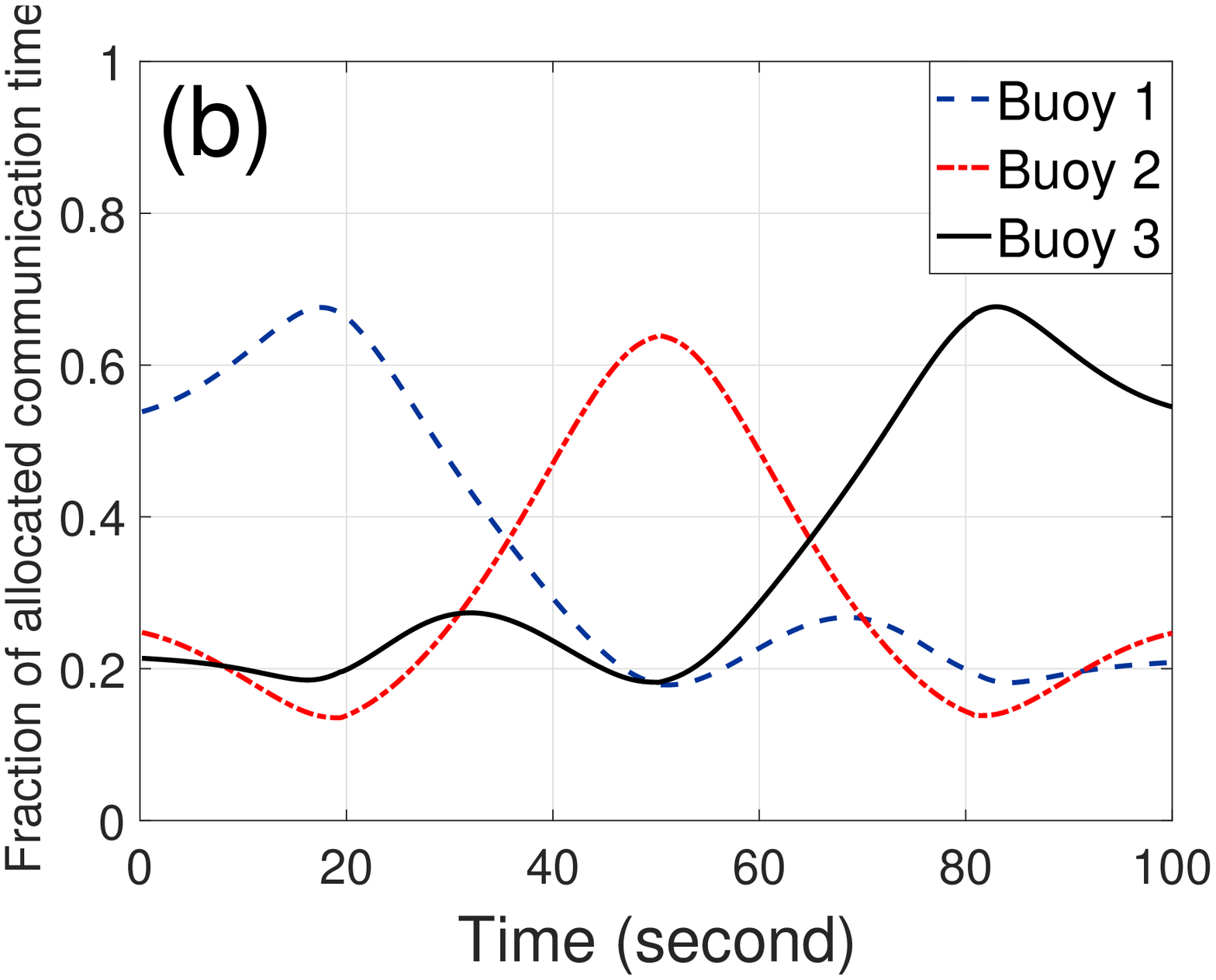}}
	\caption{(a) Optimized UAV trajectory with/without wind. (b) Communication time allocation, with wind (similar for the case without wind).\vspace{-1ex}}
	\label{Multi}
\end{figure}

If the data volume $Q$ is small/moderate, we could jointly optimize the trajectory and communication to reduce the UAV's energy consumption, subject to wind effect.
For illustration, consider $\mathbf{q}_{0}={[0,0]^T}$ m, $\mathbf{q}_{F}={[1200,0]^T}$ m, $V_w=10$ m/s (headwind, from east to west), and three buoy locations shown in Fig. \ref{Multi}(a).
Assume that each buoy has the same required $Q=200$ Mbits and the UAV needs to complete the task in time $T=100$ s. 
The optimized UAV trajectory and communication time allocation are shown in Fig. \ref{Multi}.
First, it can be seen from Fig. \ref{Multi}(b) that more time is allocated to a buoy when the UAV flies closer to it, which conforms to the cyclical TDMA principle to exploit the good channel associated with short UAV-buoy distance.
Second, under the headwind condition, by properly optimizing the UAV's trajectory and airspeed, the UAV can have more time to communicate with each buoy at a shorter distance, which helps to shorten the flight trajectory and thus reduce energy consumption (e.g., 11.17 kJ without wind and 10.44 kJ with wind).
This thus validates the effectiveness of our joint trajectory and communication design in minimizing the UAV's energy consumption subject to wind effect.

\subsubsection{Short Buoy Distance}
\begin{figure}[h]
	\centering
	\includegraphics[height=0.75\linewidth,width=0.95\linewidth]{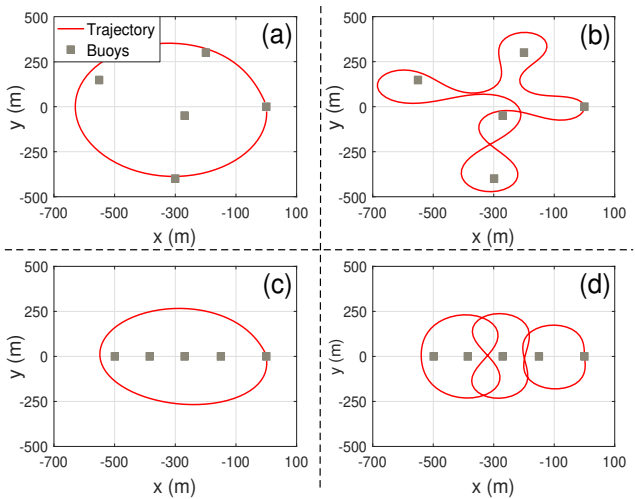}
	\caption{The optimized UAV trajectory for the topology with spread buoys and (a) $Q_0=100$ Mbits or (b) $Q_0=200$ Mbits, or for the line topology with (c) $Q_0=100$ Mbits or (d) $Q_0=200$ Mbits.\vspace{-2ex}}
	\label{MB}
\end{figure}

In this case, the buoys are clustered in a small region.
We can then design cyclical trajectories to fit the buoys' topology and also cater for the wind effect.
Since the wind effect has been extensively discussed in the above simulation results, here we assume zero wind and hence focus on the impact of the buoys' topology and the required data volume.
For illustration, consider two topologies each with five buoy locations shown in Fig. \ref{MB}(a) and (c), respectively.
For the case with large data volume to be collected, we consider two kinds of data partitions with $Q_0=100$ Mbits and $Q_0=200$ Mbits in each lap, respectively.
The optimized UAV trajectories for different buoy topology and $Q_0$ are shown in Fig. \ref{MB}.

For the topology in Fig. \ref{MB}(a) and (b) with spread buoys,
as $Q_0$ increases, it is observed that the optimized trajectory gets closer to each of the buoys in order to collect data at a higher rate.
Similar result is observed for the line topology in Fig. \ref{MB}(c) and (d).
In addition, by comparing Fig. \ref{MB}(a) and (c) under the same $Q_0$, it is observed that the optimized trajectory becomes flat in Fig. \ref{MB}(c), which tends to fit the buoys' topology and get closer to the buoys.
The above results further validate our proposed joint trajectory and communication design in adapting to the buoys' topology under different data volume requirement.

\section{Conclusions}\label{SectionConclusion}
This paper investigates a maritime data collection system with a fixed-wing UAV dispatched as a mobile data collector, and aims to minimize its energy consumption by joint trajectory and communications optimization, subject to marine wind effect.
This problem is non-convex and difficult to solve, especially when the targeted data volume is large.
We propose a new cyclical trajectory design framework that can handle arbitrary data volume efficiently subject to wind effect, which also reduces the trajectory/computational complexity. Numerical results show that our proposed framework achieves significant energy savings compared with the benchmark one-flight-only scheme. Moreover, it is shown that our optimized cyclical trajectory is able to proactively utilize the wind to complete the data collection task more efficiently with lower energy consumption.
Finally, more cyclical trajectory patterns can be explored in future work.

\bibliography{IEEEabrv,BibDIRP}
	
\newpage
\ifCLASSOPTIONcaptionsoff
\newpage
\fi
	
\end{document}